\documentclass[%
 reprint,
superscriptaddress,
 amsmath,amssymb,
 aps,
 pra,
]{revtex4-2}

\usepackage{natbib}
\usepackage{placeins}

\AtBeginDocument{%
  \setlength{\parskip}{0.5\baselineskip}%
  \setlength{\parindent}{1em}%
}
\usepackage{graphicx}
\usepackage{dcolumn}
\usepackage{bm}
\usepackage[normalem]{ulem}
\AtBeginDocument{%
  \renewcommand*{\ul}[1]{#1}%
  \renewcommand*{\uline}[1]{#1}%
}
\usepackage[english]{babel}
\usepackage{soul}

\usepackage{color}
\usepackage[hidelinks]{hyperref}
\usepackage{xr-hyper}

\soulregister\cite7
\soulregister\ref7
\begin{document}
\raggedbottom
\preprint{APS/123-QED}

\title{Harnessing thermo-optic dynamics for frequency-agile soliton microcombs}

\author{Yang Liu}
 \altaffiliation{These authors contributed equally.}
\affiliation{DTU Electro, Department of Electrical and Photonics Engineering, Technical University of Denmark, 2800 Kongens Lyngby, Denmark}

\author{Suwan Sun}
 \altaffiliation{These authors contributed equally.}
\affiliation{Key Laboratory of Specialty Fiber Optics and Optical Access Networks, Joint International Research Laboratory of Specialty Fiber Optics and Advanced Communication, Shanghai University, Shanghai 200044, China}

\author{Yueguang Zhou}
 \altaffiliation{These authors contributed equally.}
\affiliation{DTU Electro, Department of Electrical and Photonics Engineering, Technical University of Denmark, 2800 Kongens Lyngby, Denmark}

\author{Yanjing Zhao}
\affiliation{DTU Electro, Department of Electrical and Photonics Engineering, Technical University of Denmark, 2800 Kongens Lyngby, Denmark}

\author{Chaochao Ye}
\affiliation{DTU Electro, Department of Electrical and Photonics Engineering, Technical University of Denmark, 2800 Kongens Lyngby, Denmark}

\author{Xinda Lu}
\affiliation{DTU Electro, Department of Electrical and Photonics Engineering, Technical University of Denmark, 2800 Kongens Lyngby, Denmark}

\author{Yi Zheng}
\affiliation{DTU Electro, Department of Electrical and Photonics Engineering, Technical University of Denmark, 2800 Kongens Lyngby, Denmark}

\author{Leif Kastuo Oxenløwe}
\affiliation{DTU Electro, Department of Electrical and Photonics Engineering, Technical University of Denmark, 2800 Kongens Lyngby, Denmark}

\author{Kresten Yvind}
\affiliation{DTU Electro, Department of Electrical and Photonics Engineering, Technical University of Denmark, 2800 Kongens Lyngby, Denmark}

\author{Hairun Guo}
 \homepage{hairun.guo@shu.edu.cn}
\affiliation{Key Laboratory of Specialty Fiber Optics and Optical Access Networks, Joint International Research Laboratory of Specialty Fiber Optics and Advanced Communication, Shanghai University, Shanghai 200044, China}

\author{Minhao Pu}%
 \email{mipu@dtu.dk}
\affiliation{DTU Electro, Department of Electrical and Photonics Engineering, Technical University of Denmark, 2800 Kongens Lyngby, Denmark}

\begin{abstract}
Dissipative Kerr soliton microcombs enable compact and scalable frequency comb sources for precision metrology, spectroscopy, communications, and coherent LiDAR, where broad and reliable frequency tuning is essential. Thermo-optic response can support thermal locking during soliton operation, enabling resonance tracking and thereby extending the tuning range, albeit modestly. However, it also induces pronounced thermal instability during soliton initiation, hindering reliable access to this extended operating regime and limiting practical deployment in applications requiring frequency agility. Here we show that strong mode coupling reshapes the effective detuning trajectory governing soliton formation, establishing a distinct operating regime in which thermo-optic response is significantly reinforced and constructively harnessed. In this regime, soliton formation proceeds without the thermal instability inherent to conventional operation, enabling robust soliton generation in material platforms previously limited by strong thermal effects. Importantly, the enhanced thermo-optic response strengthens thermal locking during soliton operation, enabling more effective resonance tracking and substantially extending the tuning range. Leveraging this regime in AlGaAs-on-insulator multimode microresonators, we demonstrate soliton generation with a tuning range approaching 100 GHz at a \uline{sustaining} pump power of 32 mW, \uline{following initiation at 50~mW.} The same mechanism further enables frequency-agile operation through direct pump-frequency tuning without auxiliary stabilization, allowing massively parallel chirped comb generation with more than 90 channels exhibiting frequency excursions exceeding 10 GHz. These results establish \ul{an effective approach for combining direct soliton access with enhanced thermo-optic resonance tuning}, enabling robust and frequency-agile integrated soliton microcombs.

\end{abstract}

\maketitle

\section{\label{sec:Intro}Introduction}

\noindent Strong coupling describes a regime in which interacting resonant modes exchange energy faster than they dissipate it, giving rise to hybridized eigenstates with distinct spectral splitting and spatial field distributions \cite{haroche_exploring_2006,mazzei_controlled_2007}. In optical microresonators, such modal hybridization forms supermodes whose resonance conditions and intracavity energy distributions can be engineered through controlled mode coupling \cite{bao_laser_2019, saha_intracavity_2021}. Beyond dispersion shaping and spectral control, \ul{mode coupling provides a means of tailoring} the interplay between resonance detuning, intracavity power, and thermo-optic feedback that governs soliton formation, stability, and spectral evolution.

Dissipative Kerr soliton microcombs \cite{herr2014temporal,Kippenberg2018} are pursued for applications requiring widely tunable and spectrally stable multi-wavelength light sources, including spectroscopy \cite{Suh2016,jacquet2009frequency,muraviev2020broadband,picque2019frequency}, microwave photonics \cite{Liu2020}, optical communications \cite{Marin-Palomo2017} and LiDAR \cite{Liu2020a,Riemensberger2020}. 
The soliton existence range (SER), defined as \uline{the pump-frequency tuning window} over which a stable soliton state is sustained, is a key parameter governing soliton tunability and frequency agility. \ul{Kerr nonlinear dynamics govern the effective pump–resonance detuning range that supports soliton operation, while the thermal evolution of the cavity resonances maps this detuning range onto the SER. This tuning window generally broadens with increasing} pump power due to Kerr nonlinear scaling \cite{Yi2015Soliton,jacobsen2025high}, and can be further extended by thermo-optic response through thermal locking \cite{nishimoto2022thermal,zhao2025thermal}. \ul{However, accessing this extended tuning window remains challenging because thermo-optic feedback can destabilize conventional soliton initiation during the blue-to-red detuning transition.} This instability becomes more pronounced \ul{with a stronger thermo-optic response} and higher pump power, hindering reliable access to the extended operating regime and limiting practical use in frequency-agile applications. 

\begin{figure*}[htb]
\includegraphics[width=\textwidth]{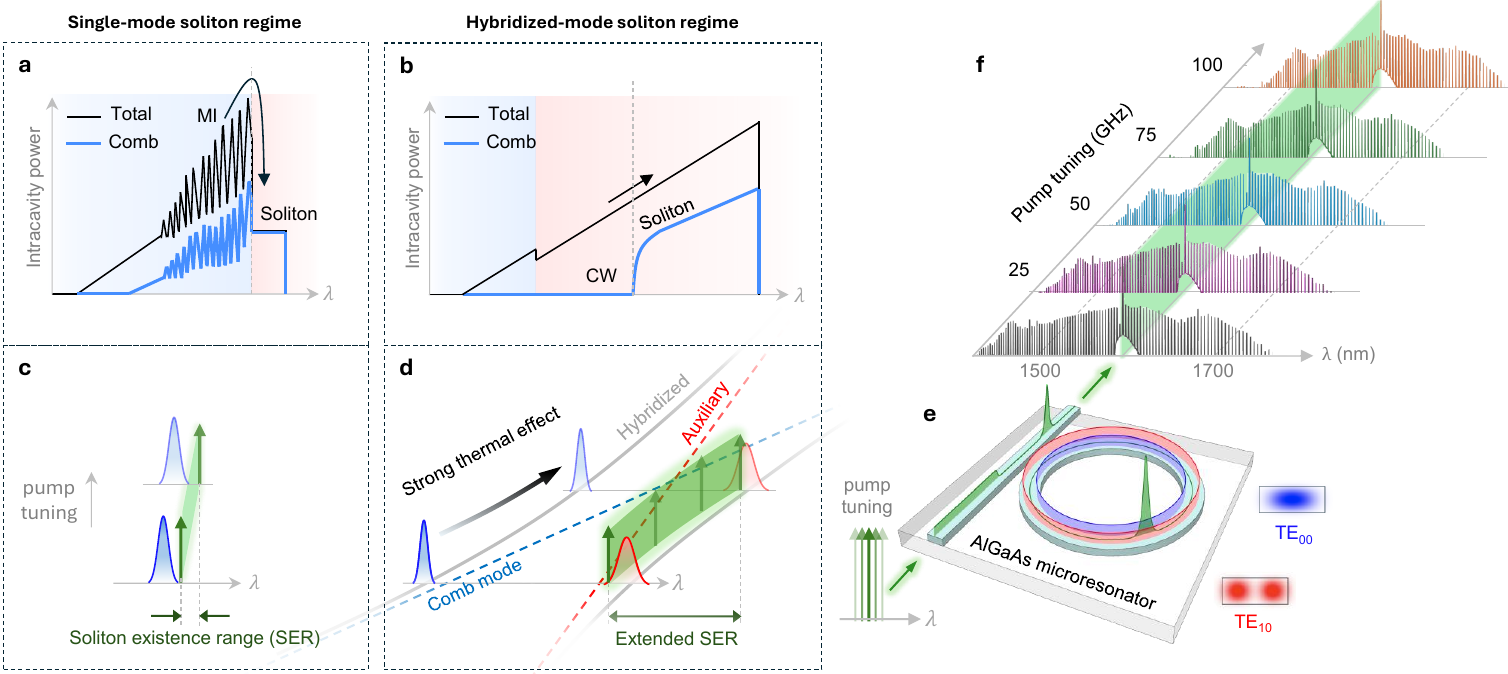}
\caption{
\textbf{Strong mode coupling enables a soliton regime harnessing thermo-optic effects.} 
\textbf{a–d}, Schematic illustration of soliton operation in the conventional single-mode regime (a, c) and the strong hybridized-mode regime (b, d). Soliton access dynamics are compared in (a) and (c), where black curves denote intracavity power and blue curves denote comb power. In (a), \ul{soliton forms from modulation instability (MI) state} at the blue-to-red detuning transition of the comb-generating resonance, where a sudden drop in intracavity power induces thermo-optic resonance walk-off and destabilizes soliton access. In contrast, in (c), strong mode coupling enables soliton formation \ul{from the continuous-wave (CW) state} with the pump coupled to a hybridized auxiliary mode on the blue-detuned side of the resonance, thereby shifting soliton formation away from the blue-to-red detuning transition and avoiding the associated thermal instability. The soliton existence ranges are compared in (b) and (d). Conventional operation supports soliton states only within a narrow red-detuned region of the comb-generating mode. In contrast, operation on the thermally stable blue-detuned branch of the hybridized auxiliary mode leverages enhanced thermo-optic feedback to extend the soliton existence range. 
\textbf{e}, Schematic of the AlGaAs multimode microresonator implementing the hybridized-mode regime. Mode hybridization between the TE$_{00}$ and TE$_{10}$ modes enables the detuning reconfiguration required for soliton formation. 
\textbf{f}, Measured soliton comb spectra as a function of pump frequency. Across the tuning range enabled by mode hybridization, the soliton state and its spectral envelope are sustained over a pump tuning range approaching 100 GHz.}
\label{fig1_overview}
\end{figure*}

This limitation becomes particularly severe in material platforms with strong Kerr nonlinearity and thermo-optic response, including AlGaAs \cite{pu2016efficient}, InGaP \cite{eckhouse2010highly}, GaP \cite{wilson2020integrated} and silicon-rich nitride \cite{kruckel2015linear}, where both effects favor a large SER. However, strong thermo-optic feedback in these platforms often prevents \ul{soliton access} through the conventional detuning pathway unless stringent dispersion and resonator design conditions are satisfied \cite{Wu2023}, \uline{posing a trade-off between SER extension and thermal accessibility of the soliton state}. \ul{Cryogenic cooling reduces thermo-optic coefficients of the material and enables soliton access in these platforms \cite{Moille2020}, but at the expense of reducing the thermal contribution to SER extension. Auxiliary-laser thermal compensation has also enabled stable soliton access while retaining a residual thermal response that extends the SER, thereby offering a means of balancing this trade-off} \cite{zhao2025thermal}.

\ul{Mode coupling offers another means of controlling soliton access and thermal stability. It} 
\ul{has been widely investigated for tailoring microcomb properties and operation, including dispersion engineering}~\cite{lucas_tailoring_2023}, \ul{control of breather dynamics}~\cite{hu_spatio-temporal_2024}, and coherent self-injection-locked operation~\cite{Ulanov2024Synthetic}. \ul{In SiN microresonators, mode coupling has enabled stable soliton generation with thermal compensation provided by an auxiliary mode whose polarization is either the same as or orthogonal to that of the comb-generating mode~\cite{zheng2026engineered, liu2025breaking, wang2024polarization}. Blue-detuned operation in these demonstrations has also enabled access to larger effective pump–resonance detunings for broader comb spectra~\cite{liu2025breaking} or an extended SER~\cite{wang2024polarization}.}

\ul{A complementary strategy is to modify the soliton access pathway itself, avoiding the abrupt intracavity-power drop that can thermally destabilize soliton formation, while retaining the thermo-optic response for SER extension. Other mode-coupling implementations~\cite{helgason2023surpassing,zhu2026power,Yu2021Spontaneous} have demonstrated modified access dynamics, but have not combined these dynamics with substantial SER extension through thermo-optic feedback.}

Here, we show that strong mode coupling reshapes the effective detuning trajectory governing soliton formation, enabling a distinct operating regime \ul{that combines direct soliton access with substantial thermo-optic extension of the pump-frequency tuning range. We realize this regime in AlGaAs-on-insulator (AlGaAsOI) microresonators, exploiting their large Kerr nonlinearity~\cite{stassen2019ultra, kim2024parity} and pronounced thermo-optic response~\cite{Moille2020}. Solitons are accessed directly from the continuous-wave (CW) state while the pump remains on the blue side of the pumped hybridized resonance. Unlike the SiN demonstrations~\cite{zheng2026engineered, liu2025breaking, wang2024polarization} discussed above, which use thermal compensation to accommodate soliton formation from intermediate comb states, this pathway avoids an abrupt intracavity-power drop in the comb-generating mode and the associated thermal destabilization. It therefore enables stable soliton access despite the strong thermo-optic response of AlGaAsOI, without requiring compensation for the initial formation transient. This same response is then harnessed through blue-detuned thermal locking, allowing the cavity resonance to track the pump over a broad frequency range. Together, the modified access pathway and the large material thermo-optic response enable frequency-agile soliton operation without auxiliary control.}

\ul{Experimentally, we demonstrate stable soliton generation in AlGaAsOI microresonators with a record soliton existence range approaching 100 GHz at a sustaining pump power of 32 mW following initiation at approximately 50 mW.} We further demonstrate a massively parallel chirped multi-frequency comb source \ul{with a wide pump-frequency excursion of 11.4 GHz}, illustrating the frequency agility enabled by this regime. \ul{These results open a route to exploiting strongly thermo-optic material platforms for widely tunable soliton microcomb sources.}

\section{\label{sec:results}Results}

\noindent\textbf{Principle of operation}

Multimode microresonators are commonly used to satisfy dispersion and low-loss requirements for soliton generation, but they inherently introduce inter-mode coupling that perturbs resonance frequencies, dispersion, and quality factors. Consequently, conventional soliton generation relies on pumping resonances of the comb-generating mode family that remain effectively unperturbed (single-mode regime; Figs.~\ref{fig1_overview}a and \ref{fig1_overview}c). Here, we instead exploit strong mode coupling to \ul{implement hybridized-mode soliton operation in AlGaAsOI.} By pumping hybridized resonances formed through inter-mode coupling between a comb-generating mode and an auxiliary mode (Figs.~\ref{fig1_overview}b and \ref{fig1_overview}d), the effective detuning condition between the pump and the comb-generating mode for initiating soliton formation is altered. As a result, soliton formation follows a modified dynamical pathway, thereby reshaping the intracavity power dynamics.

Figs.~\ref{fig1_overview}a and \ref{fig1_overview}b illustrate the intracavity power evolution during soliton formation in the single-mode and hybridized-mode regimes. In conventional operation, the pump is tuned across a single-mode resonance from the blue- to the red-detuned regime to access the soliton state. This transition is accompanied by an abrupt drop in intracavity power (Fig.~\ref{fig1_overview}a), inducing a rapid thermally induced resonance walk-off from the pump and destabilizing soliton access. Previous approaches mitigate thermal instability through techniques such as fast tuning, auxiliary fields, and feedback control \cite{herr2014temporal,Zhang2019sub,Zhou2019,Brasch2016,Weng2021,pavlov2018narrow,shen2020integrated,obrzud2017temporal,li2017stably}. In contrast, when pumping a hybridized resonance, soliton formation occurs on the blue side of the long-wavelength hybridized resonance. In this configuration, the pump remains effectively red-detuned relative to the comb-generating mode, and coupling through the hybridized resonance enables efficient intracavity power build-up for the comb-generating mode. As a result, soliton formation is displaced from the conventional blue-to-red detuning transition and proceeds with a smooth intracavity power evolution (Fig.~\ref{fig1_overview}b), thereby avoiding the associated thermal instability.

This modification of the access dynamics directly impacts the achievable SER. For materials with a positive thermo-optic coefficient, a positive intracavity power gradient during pump red-tuning causes the resonance to follow the pump, thereby extending the SER. \ul{In the single-mode pump scheme illustrated in Fig.~\ref{fig1_overview}c}, the pump is far red-detuned, resulting in inefficient coupling and intracavity power dominated by the weak soliton component. Consequently, the thermo-optic response remains limited, providing only modest SER extension. In contrast, operation on the blue side of the hybridized resonance enables efficient near-resonant coupling, producing a pronounced increase in intracavity power during red-tuning and thus strong thermo-optic locking, resulting in a significantly extended SER (Fig.~\ref{fig1_overview}d). This demonstrates that the thermo-optic response can be effectively utilized, enabling robust soliton operation in platforms with a strong thermo-optic effect.

We implement this hybridized-mode pumping strategy in an AlGaAs multimode microresonator \cite{Ye2024TMultimode} (Fig.~\ref{fig1_overview}e). The $\mathrm{TE_{00}}$ mode serves as the comb-generating mode, while the $\mathrm{TE_{10}}$ mode acts as the auxiliary mode. The stronger optical confinement of the $\mathrm{TE_{00}}$ mode results in a larger thermo-optic shift, providing the differential thermal response required for operation in the hybridized-mode regime. Enabled by this effect, we experimentally demonstrate stable single-soliton operation with a pump tuning range approaching 100~GHz (Fig.~\ref{fig1_overview}f). This regime also enables stable spectral characteristics across the tuning range. These results establish a soliton operating regime in which thermally stable access and enhanced thermo-optic tuning are achieved simultaneously. This combination is particularly important for frequency-agile operation, as it enables continuous and controlled frequency tuning while maintaining stable soliton states, as examined in detail in the following sections.\\

\begin{figure*}[ht]
\includegraphics[width=\textwidth]{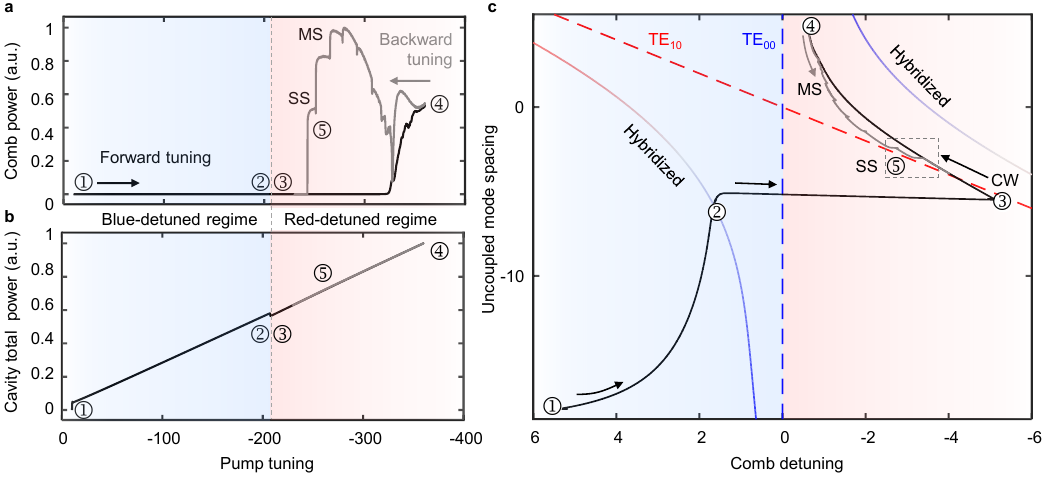}
\caption{\textbf{Detuning trajectory of soliton generation in the hybridized-mode regime.} \textbf{a, b}, Evolution of comb power (a) and total intracavity power (b) during pump tuning across a pair of strongly coupled spatial modes. Arrows indicate the tuning direction. Forward tuning leads to a transition from the continuous-wave (CW) state to a multi-soliton (MS) state, while backward tuning converts the MS state into a single-soliton (SS) state. \textbf{c}, Calculated pump detuning trajectory relative to the hybridized resonances. Dashed red and blue lines denote the intrinsic (uncoupled) $\mathrm{TE_{10}}$ and $\mathrm{TE_{00}}$ resonances, while solid curves represent the hybridized (coupled) modes formed through strong mode coupling. The pump tuning in (a, b) and comb detuning in (c) are normalized to the linewidth of the $\mathrm{TE_{00}}$ comb-generating mode.
}
\label{fig_simulation}
\end{figure*}

\noindent\textbf{Simulated soliton dynamics in hybridized resonances}

To further \uline{qualitatively} investigate the soliton formation dynamics \uline{under slow
pump tuning} in the hybridized configuration, we numerically solve a set of coupled Lugiato–Lefever equations (LLEs) that capture the nonlinear evolution and inter-mode coupling between the primary comb mode and the auxiliary mode (see Methods). Figs.~\ref{fig_simulation}a,b present the simulated evolution of the comb power and the total power in the microresonator, respectively, as the pump frequency is tuned across the hybridized resonances. Fig.~\ref{fig_simulation}c tracks the evolution of the comb detuning, providing direct insight into the detuning trajectory enabling smooth soliton formation. In Fig.~\ref{fig_simulation}c, the dashed curves denote the spectral positions of the $\mathrm{TE_{00}}$ and $\mathrm{TE_{10}}$ resonances in the absence of coupling, while the solid curves represent the positions of the hybridized $\mathrm{TE_{00}}$ and $\mathrm{TE_{10}}$ resonances, referred to as $\mathrm{TE_{00}}$-like resonance and $\mathrm{TE_{10}}$-like resonance. The avoided mode crossing (AMX) arises from the inter-mode coupling.

The operation begins with forward pump tuning from state 1 to state 4. As the pump approaches the $\mathrm{TE_{00}}$-like resonance, the intracavity power increases, leading to a rise in temperature. Because the $\mathrm{TE_{00}}$ mode exhibits a larger thermo-optic shift than the $\mathrm{TE_{10}}$ mode, the spectral spacing between the two hybridized resonances decreases from state 1 to state 2. When the pump traverses the $\mathrm{TE_{00}}$-like resonance at state 2, the CW state is maintained, and the pump undergoes a transition from blue to red detuning, accompanied by a rapid decrease in intracavity power and temperature. As the pump subsequently enters the hybridized $\mathrm{TE_{10}}$-like resonance on its blue side at state 3, the intracavity power is re-established. Multi-soliton generation occurs during the subsequent forward tuning from state 3 to state 4, where the pump remains on the blue side within the hybridized resonance. \ul{This process is crucial for the system to access soliton bifurcation, which is also revealed to be a smooth and adiabatic transition, as shown in Fig.~\ref{fig_simulation}b.}

A subsequent backward tuning toward state 5 reveals multiple small steps in the comb power, corresponding to transitions from higher-number to lower-number soliton states and ultimately to a single soliton. These discrete soliton switching events are consistent with simulations reported in other coupled systems \cite{helgason2023surpassing}. Crucially, the total intracavity power in Fig.~\ref{fig_simulation}b remains continuous and smooth \ul{during these soliton transitions}. The power variations in the $\mathrm{TE_{00}}$ mode are compensated by the $\mathrm{TE_{10}}$ mode, preventing abrupt power collapse. Consequently, all soliton states reside on a continuous positive power slope, enabling adiabatic evolution toward a single-soliton state. Moreover, the steep intracavity power slope in Fig.~\ref{fig_simulation}b enables the resonance to track the pump tuning over an extended range. The single-soliton state at state 5 also occupies a broad thermally stable region rather than a narrow operating point. A broadened and continuous stability region is essential for frequency-agile operation, as it enables reliable and uninterrupted frequency tuning over a wide range. \ul{Additional simulations examining the dependence
of soliton access on dispersion and intermodal coupling
are provided in the Supplementary Information. These simulations provide a qualitative physical picture
of the soliton-access pathway, which is investigated
experimentally below.}\\

\begin{figure*}
\includegraphics[width=\textwidth]{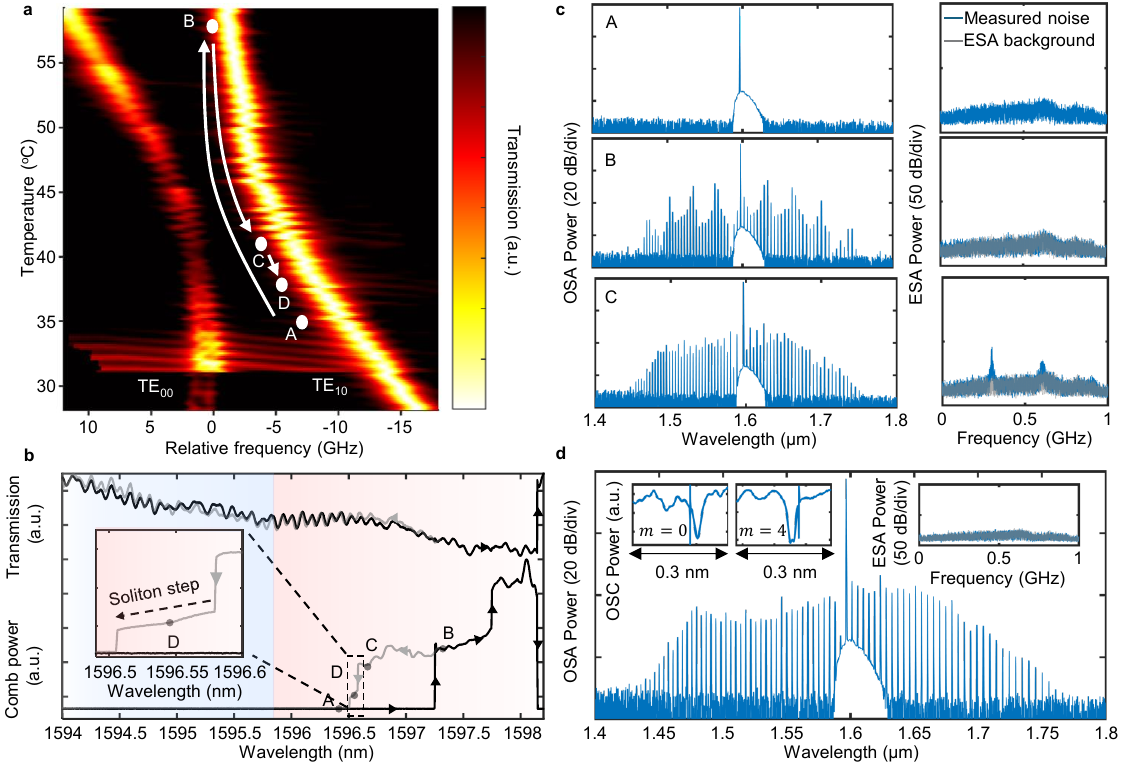}
 
\caption{\textbf{Experimental dynamics of soliton access in a multimode microresonator.} \textbf{a}, Transmission map at different temperatures showing the relative detuning between the $\mathrm{TE_{00}}$ and $\mathrm{TE_{10}}$ resonances. As the two modes approach each other, an avoided-mode crossing (AMX) forms. Arrows indicate the pump trajectory during single-soliton access.  \textbf{b}, Cavity transmission (top) and comb power (bottom) during forward (black) and backward (gray) pump scans across the hybridized resonances. The system evolves from the CW state (A) to a MS state (B) during forward tuning, and transitions through a breathing-soliton state (C) to a SS state (D) during backward tuning. \textbf{Inset:} zoom of the soliton step. \textbf{c}, Optical spectra and corresponding RF noise spectra measured at states A–C. \textbf{d}, Optical spectrum of the SS state (D). Left inset: detuning measurement for pump and a comb line (m = 4). Right inset: RF noise spectrum confirming the low-noise SS state.}
\label{fig2_accessDynamics}
\end{figure*}

\noindent\textbf{Thermally stable soliton access in AlGaAs microresonators}

Our fabricated AlGaAsOI soliton devices consist of multimode microresonators with cross-sectional dimensions of $\mathrm{380\ nm\times750 \ nm}$. The devices are patterned on AlGaAsOI wafers \cite{ottaviano2016low} using electron-beam lithography, followed by dry etching \cite{zheng2019high,kim2022design}. The inset of Fig.~\ref{fig1_overview}e shows the mode profiles of the supported TE$\mathrm{_{00}}$ and TE$\mathrm{_{10}}$ modes. The multimode microresonator is coupled to a single-mode bus waveguide with a point coupler. The FSR of the TE$\mathrm{_{00}}$ mode is 520 GHz, the second-order dispersion $D_2/2\pi$ is 0.22 MHz, and the third-order dispersion $D_3/2\pi$ is -1.46 MHz. The linewidth of the $\mathrm{TE_{00}}$ mode is measured to be around 1.2 GHz. The TE$\mathrm{_{00}}$ and TE$\mathrm{_{10}}$ resonances exhibit thermal shifts of 0.106 $\mathrm{nm/^{\circ}C}$ and 0.099 $\mathrm{nm/^{\circ}C}$, respectively.

Fig.~\ref{fig2_accessDynamics}a shows the relative frequency shift of the hybridized resonances from the uncoupled resonance frequency of the TE$\mathrm{_{00}}$ mode at various temperatures. At room temperature, the TE$\mathrm{_{10}}$-like resonance is significantly red-detuned from the TE$\mathrm{_{00}}$-like resonance, with a separation of 17 GHz. As the temperature increases, the TE$\mathrm{_{00}}$-like resonance approaches the TE$\mathrm{_{10}}$-like resonance due to their different thermal shift rates. A significant AMX is observed at around 45 $\mathrm{^{\circ}C}$, inducing a strong frequency shift and hybridization of both resonances. The coupling rate between the two modes, extracted from the minimum separation of the hybridized resonances, is approximately 7.5 GHz.

\begin{figure*}
\includegraphics[width=1\linewidth]{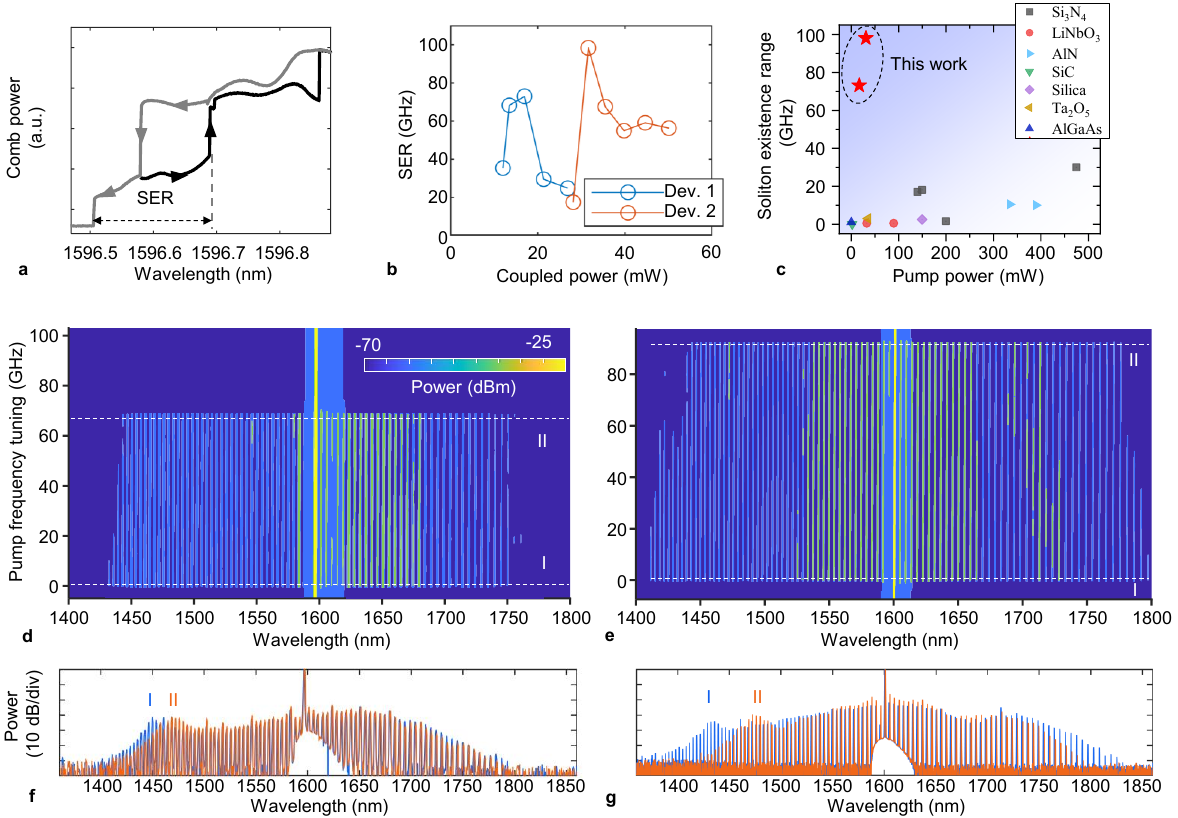}
\caption{\textbf{Large soliton existence range and robust spectra in an AlGaAs microresonator.}  \textbf{a}, Measured comb power evolution during pump tuning. The black curve corresponds to forward tuning of the pump, while the gray curve corresponds to backward tuning. \uline{The SER denotes the pump-frequency window over which
the single-soliton state is sustained as the cavity
resonances evolve thermally.} \textbf{b}, Measured SER as a function of coupled pump power for two devices (device 1 and device 2). A maximum SER of 98 GHz is obtained at a \uline{sustaining} pump power of 32 mW, \uline{following soliton
initiation at approximately 50~mW.} \textbf{c}, \uline{Comparison of experimentally reported pump-frequency
SERs under thermally evolving resonance conditions as a
function of on-chip pump power. Power from separate
auxiliary lasers is excluded, whereas power in
modulation-generated auxiliary sidebands is included.} 
\uline{\textbf{d,e}, Evolution of soliton comb spectra at sampled
pump wavelengths within the SER for device 1 (d) and
device 2 (e), showing only minor variation in comb bandwidth. 
The recorded spectra cover pump-frequency spans of around
62~GHz and 92.5~GHz, respectively. The pump powers used for devices 1 and 2 are 13.5~$\mathrm{mW}$, and 32~$\mathrm{mW}$, respectively.}
\textbf{f,g}, Representative comb spectra at the smallest (I) and largest (II) \uline{sampled} pump wavelengths within the SER for device 1 (f) and device 2 (g). The comb bandwidth does not follow the conventional monotonic increase with pump red-tuning, reflecting the modified mapping between pump wavelength and effective comb detuning in the hybridized-mode regime.}
\label{fig3_SER} 
\end{figure*}

We pump the device with 27~$\mathrm{mW}$ of coupled pump power.
Fig.~\ref{fig2_accessDynamics}b shows the corresponding transmission spectrum (upper curves) and the converted comb power (lower curves) during forward tuning. Figs.~\ref{fig2_accessDynamics}c and \ref{fig2_accessDynamics}d show the characteristic comb states during the tuning process and the corresponding noise measurements. The pump detuning relative to the hybridized resonances is monitored during the process and indicated in Fig.~\ref{fig2_accessDynamics}a. We first tune the pump forward toward the TE$\mathrm{_{00}}$ resonance from the blue side (black curves in Fig.~\ref{fig2_accessDynamics}b), causing significant cavity heating.
The temperature increase induced by pump tuning moves the two resonances closer to the strong-hybridization region. When the pump crosses the $\mathrm{TE_{00}}$-like resonance and enters the $\mathrm{TE_{10}}$-like resonance, only the CW state is observed (state A). Subsequent forward tuning further increases the intracavity power until a multi-soliton state (state B) is directly generated from the CW state. Further forward tuning leads to another multi-soliton state and eventually moves the pump off resonance. The generation of the multi-soliton state is thermally stable because it occurs while the pump remains within the hybridized resonance on its blue side.

\begin{figure*}[htpb]
 \includegraphics[width=\textwidth]{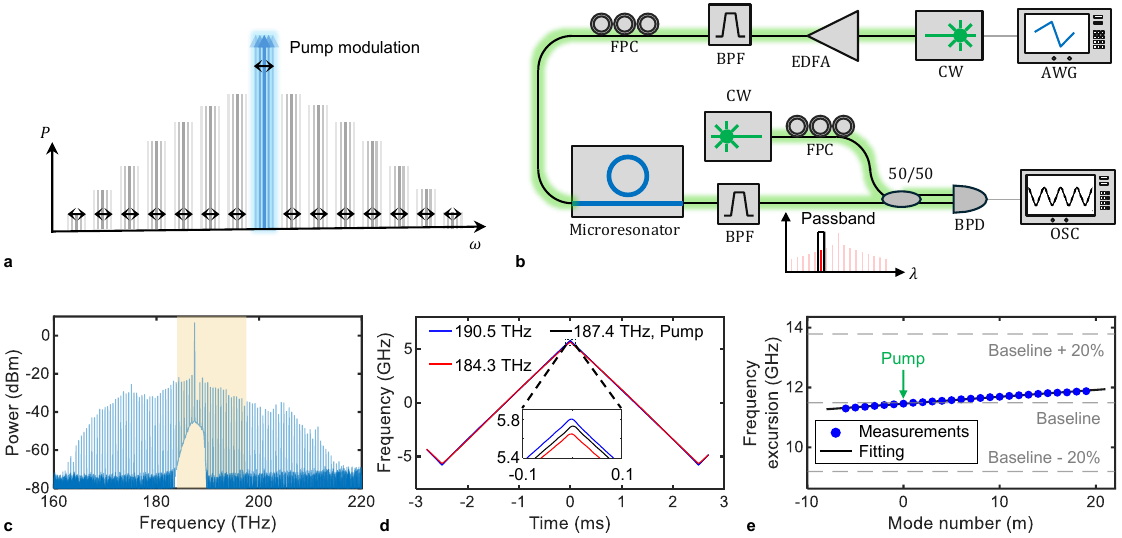}
 \caption{\textbf{Massively parallel chirped soliton comb. }\textbf{a}, Schematic of pump chirped soliton comb. When the pump frequency is modulated, all the comb lines remain phase coherent and follow the modulation, enabling parallel frequency chirping across the comb spectrum. \textbf{b},  Experimental setup for characterizing the chirped comb. AWG: arbitrary waveform generator; \ul{CW: continuous-wave laser}; EDFA: erbium-doped fiber amplifier; \ul{BPF}: band-pass filter; \ul{FPC: fiber polarization controller}; BPD: balanced photodiode; OSC: oscilloscope. \textbf{c}, Measured optical spectrum of the soliton comb used for chirping. The 25 comb lines characterized in the measurement are highlighted in the yellow-shaded region. \textbf{d}, Measured frequency shifts of the pump and two representative comb lines as a function of time for a pump chirp range of \ul{11.4 GHz}. The comb lines follow the pump modulation coherently. \textbf{e}, Measured frequency excursion of each comb line, showing minimal variation across 25 comb lines and indicating uniform frequency tracking under pump chirping.} 
 \label{fig4_chirp} 
 \end{figure*}
 
To generate the single-soliton state, we reverse the tuning direction when a multi-soliton state (state B) appears. During backward tuning (gray curves in Fig.~\ref{fig2_accessDynamics}b), a breather state (state C) is generated. Further backward tuning generates a single-soliton state (state D), with a sudden change in the comb-power trace. \uline{In our experiment, the single-soliton state can be accessed repeatably by performing backward tuning from the multi-soliton state.} A significant dispersive wave is observed at around 1480 nm and contributes to a pronounced red shift (spectral recoil) of the soliton spectral peak \cite{cherenkov2017dissipative}. The inset of Fig.~\ref{fig2_accessDynamics}d and Extended Data Fig.~\ref{ExFig2_combpumpdetuning} show the detuning measurement results performed in the single-soliton state. The comb lines are red-detuned from their respective resonances, as expected for a typical soliton comb. By contrast, the pump lies within the hybridized resonance on its blue side, in line with our numerical investigation. The pump remains on the blue-detuned side of the hybridized resonance throughout the process of multi-soliton formation and transition to the single-soliton state, providing a thermally stable pathway for soliton generation. \ul{We also observed that the minimum single-soliton initiation power is
strongly correlated with the initial mode spacing. Moreover, a similar single-soliton access pathway is also
observed in an AlGaAsOI microresonator with a
different waveguide cross section.} Overall, the experimentally observed thermally stable access dynamics establish a robust platform for continuous and controlled frequency tuning via direct pump modulation, which is a key requirement for frequency-agile microcomb operation.\\

\noindent\textbf{Enhanced soliton existence range in the hybridized-mode regime}

After soliton formation, we further characterize the SER of the single-soliton state. Fig.~\ref{fig3_SER}a shows that the backward tuning (gray curve) from the multi-soliton state leads to a clear single-soliton step. We found that an additional forward tuning (black curve) from the single-soliton step can reveal the single-soliton step that is hidden under the multi-soliton state due to the partial degeneracy between multi-soliton and single-soliton states \cite{Guo2017}. \uline{Therefore, the SER is determined from the stability boundaries
measured in two separate scans in opposite directions
from a selected single-soliton operating point}. Interestingly, the single DKS state can transition back to the previous multi-soliton state during this additional forward tuning.

Fig.~\ref{fig3_SER}b shows the SER as a function of pump power for two soliton devices. The lowest pump powers required to initiate the single-soliton state are 22 mW for device 1 and 50 mW for device 2, respectively. We found that the single-soliton state can be sustained at a lower power than the above-mentioned soliton initiation power. In both devices, the SER gradually increases and then vanishes as the pump power decreases. \uline{A maximum SER of 98~GHz is obtained in device~2
after initiating the single-soliton state at approximately
50~mW and reducing the pump power to 32~mW
for sustained operation.} These large SERs are attributed to the substantial intracavity power slope of the soliton step introduced by operating the pump on the blue side of the resonance and the large thermo-optic coefficient of AlGaAs. Fig. \ref{fig3_SER}c compares the SER across different material platforms and pump powers \ul{\cite{Wu2023,
Weng2021,liu2021aluminum,gong2019soliton,He2019Self,
li2017stably,nishimoto2022thermal,Weng2022,Weng2024,
guidry2022quantum,Yu2021Spontaneous,zhao2025thermal}}. Our results—marked by dashed circles—were achieved with remarkably low pump power enabled by AlGaAs’s strong material nonlinearity, and the wide tunability provided by its pronounced thermal response together with a blue-detuned pump.

In a conventional pumping scheme with the pump located on the red-detuned side of the comb mode, forward tuning the pump in a soliton state leads to a significant increase of comb detuning, resulting in pronounced variations in soliton bandwidth and power \cite{Yi2015Soliton,Lucas2017}.
Such a dramatic bandwidth change reduces the useful tuning range of the soliton state. By contrast, in this work the pump is located on the $\mathrm{TE_{10}}$-like resonance. The strong thermo-optic effect and the different thermal shift rates of the $\mathrm{TE_{00}}$ and $\mathrm{TE_{10}}$ modes significantly modify the conventional relation between pump tuning and comb detuning. Figs.~\ref{fig3_SER}d and \ref{fig3_SER}e show the spectral evolution of the soliton state during forward tuning for device~1 at 13.5~mW and device~2 at 32~mW, respectively. Figs.~\ref{fig3_SER}f and \ref{fig3_SER}g show representative soliton spectra at the shortest \uline{sampled} pump wavelength (I) and the longest \uline{sampled} pump wavelength (II) for device~1 and device~2, respectively. Forward tuning along the soliton steps (from state I to II) in both devices results in only a slight decrease in soliton bandwidth. This suggests that the effective comb detuning decreases during forward tuning in our configuration, consistent with the simulated detuning trajectory of the single-soliton state in Fig.~\ref{fig_simulation}c. Moreover, device~1 exhibits a much smaller bandwidth variation than device~2, highlighting the potential of using the different thermal shift rates to control the correlation between pump tuning and comb detuning. Finally, the stable comb spectrum over a wide pump-tuning range enables a large useful SER and provides a wide and continuous frequency tuning window that directly supports frequency-agile operation needed for practical applications \cite{nishimoto2022thermal}. 

\noindent\textbf{Massively parallel widely chirped comb source}

In conventional microcomb systems, achieving wide and stable frequency tuning typically requires simultaneous control of both the pump frequency and the cavity resonance through thermal, mechanical, or auxiliary-laser-based feedback \cite{liu2020monolithic,fujii2023versatile,fujii2024mechanically}. \uline{Such co-tuning schemes require additional actuators or
coordinated control.} In contrast, the present regime enables direct pump-frequency tuning without auxiliary control, as the cavity resonance intrinsically follows the pump via thermo-optic feedback. This feature, combined with the large SER, significantly simplifies system operation and enables \uline{wide-range,} continuous frequency chirping, a direct manifestation of frequency-agile operation, which is essential in high-resolution FMCW LiDAR applications. 

Fig.~\ref{fig4_chirp}a shows a schematic of a widely chirped comb source, and Fig.~\ref{fig4_chirp}b shows the setup used to generate and characterize the chirped comb (see Methods). Fig.~\ref{fig4_chirp}c shows the soliton spectrum in device~2 used for the chirping experiment. The pump is chirped across 11.4~GHz with a triangular waveform at a scan rate of 200~Hz. More than 90 comb channels across the spectrum exhibit a small power variation of 0.22~dB on average during chirping. We characterize the chirping of the highlighted 25 comb lines. The triangular waveform of the pump chirp is faithfully transferred to the comb lines, as shown in Fig.~\ref{fig4_chirp}d and Extended Data Fig.~\ref{ExFig4_frequency-time-map}. As the repetition rate of the comb changes during pump tuning, the frequency excursions of the comb lines are not identical, which is a common feature of this approach. Since the comb detuning is relatively stable during pump tuning in our case, the repetition rate shifts by only 2~MHz per gigahertz of laser tuning. This small repetition-rate deviation allows more useful comb teeth with a uniform and wide frequency-excursion range. The minor mismatch in frequency excursion across the comb lines can be easily calibrated in post-processing. Consequently, all comb lines in the spectrum chirp over more than 10~GHz and can potentially be used in an FMCW LiDAR system, corresponding to centimeter to sub-centimeter ranging resolution—about an order of magnitude better than other soliton-microcomb chirping demonstrations \cite{Liu2020a,Riemensberger2020}.

\section{\label{sec:level1}Discussion}

\ul{We have demonstrated thermally stable soliton access and wide-range pump-frequency tuning in AlGaAsOI
microresonators by reconfiguring the soliton access dynamics through strong spatial-mode coupling.
This permits direct CW-to-soliton access on the blue
side of the pumped hybridized resonance, with smooth
total intracavity-power evolution that avoids the
thermal instability associated with conventional
soliton initiation. Together with thermo-optic resonance tracking during blue-detuned operation, this access pathway enables a soliton existence range approaching 100 GHz at a sustaining pump power of 32 mW following initiation at approximately 50 mW.} The modest pump power lies within the range accessible by integrated laser sources, highlighting the potential for fully integrated microcomb systems.

The ability to operate through direct pump-frequency tuning without auxiliary resonance control provides a simplified and scalable route to frequency-agile microcomb operation. \uline{This intrinsic tracking between the pump and cavity resonance
supports continuous tuning over a wide frequency range while
maintaining stable soliton states. The measured small-signal
3-dB resonance-response bandwidth of approximately 70~kHz
suggests potential for pump frequency tuning at modulation frequencies above the demonstrated 200 Hz. Operation at modulation frequencies approaching or even exceeding the measured 70 kHz bandwidth may be possible with tuning excursions smaller than the full SER, as thermal lag limits resonance tracking. Such operation above the thermal response bandwidth has been demonstrated in a related thermal-control scheme, albeit with reduced excursions~\cite{nishimoto2022thermal}. The large SER under slow tuning may provide additional margin for maintaining useful frequency excursions at higher modulation frequencies.}\\

Beyond stable access and tunability, the observed \ul{reversal of the comb detuning response to pump-frequency tuning} reveals that thermo-optic effects can act as an effective control knob for soliton dynamics. This behavior enables regulation of comb detuning over a wide tuning range, in contrast to conventional systems where detuning varies strongly with pump tuning and leads to pronounced phase-noise transduction \cite{Yang2021,Lei2022}. Maintaining stable or controlled comb detuning therefore provides a pathway toward low-noise soliton operation.

As a multi-wavelength chirped optical source, soliton microcombs can offer broader optical bandwidth and a larger number of usable channels compared to dark-pulse microcombs \cite{Shu2023}\ul{, making it attractive} for high-resolution parallel FMCW LiDAR. \ul{Spatial mode coupling may further extend the optical bandwidth of the soliton state \cite{liu2025breaking}, potentially enabling an even larger number of usable comb lines for massively parallel ranging.} With further optimization of dispersion engineering and coupling conditions to enhance comb-line power, this platform provides a promising route toward practical frequency-agile integrated photonic sources. \ul{More broadly, this work illustrates how modifying soliton access dynamics can enable constructive use of strong thermo-optic feedback, offering a route to widely tunable microcombs in material platforms where thermal instability otherwise restricts operation.} \\

\noindent\textbf{Acknowledgements}\\
This work is supported by European Research Council (REFOCUS 853522, ELASTIC 101213805), European Innovation Council (CSOC 101047289), Danish National Research Foundation (SPOC ref. DNRF123), Independent Research Fund Denmark (ifGREEN 3164-00307B), and Innovationsfonden (GreenCOM 2079-00040B).\\
\\
\textbf{Author contributions}\\
Y.L. and M.P. conceived the idea. Y.Zhou and C.Y. designed the device. Y.Zhou and C.Y. fabricated AlGaAsOI samples. Y.L., Y.Zhou, Y.Zhao, C.Y., and X.L. conducted the experiment with assistance from Y.Zheng. Y.L., S.S., and H.G. contributed to the theoretical understanding. Y.L., S.S., Y.Zhou, Y.Zhao, C.Y., L.K.O., K.Y., H.G., and M.P. analyzed the data. L.K.O., K.Y., H.G., and M.P. supervised the project. Y.L., S.S., H.G., and M.P. wrote the manuscript with input from all other authors.\\
\\
\textbf{Competing interests}\\
M.P., K.Y. and Y.L. are listed as inventors on a patent application that broadly covers aspects related to this work. The remaining authors declare no competing interests.\\
\\
\textbf{Data availability}\\
The data supporting the findings of this study are available from the corresponding authors upon reasonable request.\\
\\


\bibliography{apssamp}

\section{\label{sec:level1-method}Method}
\subsection{\label{sec:level2}Characterization of thermal shift in AlGaAs microresonator}

The thermal shift rates of the two modes are measured by capturing the transmission spectrum at several stage temperatures. To quickly characterize the full dynamic behavior of the coupled resonance over a wide temperature range, as shown in Fig.~\ref{fig2_accessDynamics}a, we direct a high-power laser to a single resonance located far from the coupled resonances. By slowly tuning the high-power laser into this distant resonance, the microresonator is gradually heated. Meanwhile, a counter-propagating low-power probe beam, as illustrated in Fig.~\ref{ExFig_CombCharacterizationSetup}, is swept across the coupled resonances to track changes in the transmission spectrum at different wavelengths of the high-power laser. Tuning the high-power laser into the resonance introduces a non-smooth (noisy) increase in temperature, likely due to a Fabry--P\'erot cavity effect caused by facet reflections. Therefore, the extracted resonance frequencies as a function of the heating-laser wavelength are first fitted with a parabolic curve; then, each measured trace is aligned to the fit to suppress the noise. Finally, the temperature axis in Fig.~\ref{fig2_accessDynamics}a is calibrated using the measured thermal shift rates of the resonances.

\subsection{\label{sec:level2b}Chirping microcomb characterization}
The pump laser used in this work is equipped with a built-in piezoelectric element for fine-tuning the laser frequency. An arbitrary waveform generator (AWG) in \ul{Fig.~\ref{fig4_chirp}b} is used to apply the control voltage to the piezoelectric element, thereby modulating the laser frequency. To achieve a linearly chirped pump, we implemented a pre-distortion scheme with the AWG to address the non-uniform frequency response of the piezoelectric element. Due to the limited bandwidth of the piezoelectric control element, we used a 200-Hz sweep rate in this proof-of-concept demonstration. \uline{A faster actuator, such as a single-sideband modulator,
could increase the pump-frequency modulation bandwidth.
Maintaining the same soliton frequency excursion at higher
modulation frequencies would also depend on the thermal response of the microresonator.}

To characterize the chirp of the comb lines, we employ a coherent detection scheme on the output side. A separate tunable laser, acting as a local oscillator, is tuned close to each comb line sequentially. The local oscillator and comb lines are mixed using a 3-dB coupler and sent to a balanced photodetector (BPD). This setup faithfully transfers the frequency chirp of the comb lines to a low-frequency chirping beat signal, which is resolved by the BPD. By applying a fast Fourier transform (FFT) to the recorded beat signal, the frequency excursion of the chirping comb lines can be extracted.

\subsection{\label{sec:level2b-modeling}LLE modeling with coupled mode equations }

We simulate the nonlinear process in the multimode microresonator through two linearly coupled Lugiato–Lefever equations (LLEs) \cite{fujii2018analysis}:
\small
\begin{multline*}
\frac{dA_1}{dt}
- i\sum_{n=2}^{\infty}\frac{i^nD_{n}}{n!}\frac{\partial^nA_1}{\partial \phi^n}
+ ig|A_1|^2A_1 = -\Bigl(\frac{K_1}{2} + i\omega_1 + i\delta T_1\Bigl) A_1\\
- i\frac{K_c}{2}A_2 - i\frac{\sqrt{K_\mathrm{ex1}K_\mathrm{ex2}}}{2}A_2+\sqrt{\frac{K_{\text{ex1}} P_\mathrm{in}}{\hbar \omega_0}}
\end{multline*}

\begin{multline*}
\frac{d\tilde{A_2}}{dt}
+ ig|\tilde{A_2}|^2\tilde{A_2} = -\Bigl(\frac{K_2}{2} + i\omega_2 + i\delta T_2\Bigl) \tilde{A_2} \\
 - i\frac{K_c}{2}\tilde{A}_{1,\mu}\delta_{\mu0} - i\frac{\sqrt{K_\mathrm{ex1}K_\mathrm{ex2}}}{2}\tilde{A}_{1,\mu}\delta_{\mu0}
+ \sqrt{\frac{K_{\text{ex2}} P_\mathrm{in}}{\hbar \omega_0}}
\end{multline*}

\begin{equation}
\frac{\partial \delta T_{\mathrm{1}}}{\partial t}
   = \frac{K_{T\mathrm{1}}}{\tau_{th}}
     \left(P_1+P_2
     \right)
     - \frac{\delta T_{\mathrm{1}}}{\tau_{th}} \notag
\end{equation}

\begin{equation}
\frac{\partial \delta T_{\mathrm{2}}}{\partial t}
   = \frac{K_{T\mathrm{2}}}{\tau_{th}}
     \left(P_1+P_2
     \right) 
     - \frac{\delta T_{\mathrm{2}}}{\tau_{th}} \notag
\end{equation}

\normalsize

where $A_1(\phi,t)$ is the slowly varying field amplitude of the comb mode in the microresonator, $\phi$ is the co-rotating angular coordinate in the microresonator, $\tilde{A_2}$ is the mode amplitude of the auxiliary mode near the pumped comb mode, $\tilde{A}_{1,\mu}$ is the mode amplitude of the $\mu$-th comb mode, $D_n$ is the $n$th-order coefficient of microresonator integrated dispersion, $K_1=K_{\mathrm{i1}}+K_{\mathrm{ex1}}$ is the total decay rate of the comb mode, including the intrinsic decay rate $K_{\mathrm{i1}}$ and external coupling decay rate $K_{\mathrm{ex1}}$, $K_2=K_{\mathrm{i2}}+K_{\mathrm{ex2}}$ is the total decay rate of the auxiliary mode, including the intrinsic decay rate $K_{\mathrm{i2}}$ and external coupling decay rate $K_{\mathrm{ex2}}$, $g$ is the nonlinear coupling coefficient, $K_c$ is the linear coupling rate between the comb mode and the auxiliary mode, $P_\mathrm{in}$ is the pump power, $\omega_1$ and $\omega_2$ are the pump detuning relative to the intrinsic resonant frequency of the comb mode and auxiliary mode, respectively, and $\delta_{\mu0}$ is the Kronecker delta. $\delta T_1$ and $\delta T_2$ are the thermal shifts of mode 1 and mode 2, respectively. $K_{T1}$ and $K_{T2}$ are the thermally induced resonance-shift coefficients for mode 1 and mode 2, respectively. $P_1$ and $P_2$ are the intracavity power of mode 1 and mode 2, respectively. $\tau_{th}$ is the thermal relaxation time. AMXs far from the pump are less relevant in this scheme, and the focus in this work is on pumping an AMX. Therefore, we introduce only one auxiliary mode that couples to one comb mode near the pump. The auxiliary mode is assumed to follow a homogeneous solution, which is valid when no comb is generated in the auxiliary spatial mode. For simplicity, we use the same nonlinear coupling rate for the comb mode and the auxiliary mode.

The comb mode used in this simulation is set with $K_{\mathrm{i1}}/2\pi = 700\ \mathrm{MHz}$ and $K_{\mathrm{ex1}}/2\pi = 350\ \mathrm{MHz}$. The auxiliary mode is set with $K_{\mathrm{i2}}/2\pi = 2500\ \mathrm{MHz}$ and $K_{\mathrm{ex2}}/2\pi = 600\ \mathrm{MHz}$. The intermodal coupling between the two modes is set to $K_{\mathrm{c}}/K_1 = 7$. The other parameters are $g = 201\ \mathrm{s^{-1}}$, $D_2/2\pi = 10\ \mathrm{MHz}$, $D_3/2\pi = -1.4\ \mathrm{MHz}$, $\mathrm{FSR} = 520\ \mathrm{GHz,\ } P_\mathrm{in} = 10 P_\mathrm{th}$. The thermal shift coefficients of the comb mode and the auxiliary mode are $K_{T1}=2250\ \mathrm{rad/(s\cdot W)}$ and $K_{T2}=2100\ \mathrm{rad/(s\cdot W)}$. $\tau_{th}$ is scaled down to mimic the slow tuning process in experiments.

The existence region of the single DKS state in Fig.~\ref{FigSI_stateMap} is probed using the following steps:
(1) A grid with a resolution of 0.01 in the normalized uncoupled mode spacing and normalized detuning is generated in the parameter space;
(2) An analytical expression of the single-soliton state is used to seed the model, and at least one grid point where the single-soliton state exists is identified;
(3) The identified grid point(s) are used as starting points in the parameter space. The neighboring grid points are then examined by propagating the model from the starting point(s) to the neighboring points within 5 photon lifetimes;
(4) After reaching a neighboring point, the simulation is continued for an additional 44.5 photon lifetimes to allow the intracavity state to stabilize;
(5) A coherence check within the final 5 photon lifetime is performed to determine the intracavity state. If a single DKS state is found, the corresponding grid point is added as a new starting point in the next iteration, and the numerical solution of the corresponding DKS state is used as the new seed.
(6) Steps (3)-(5) are repeated until no new single-soliton states are discovered.

\setcounter{figure}{0}
\begin{figure*}[b]
\includegraphics[width=\textwidth]{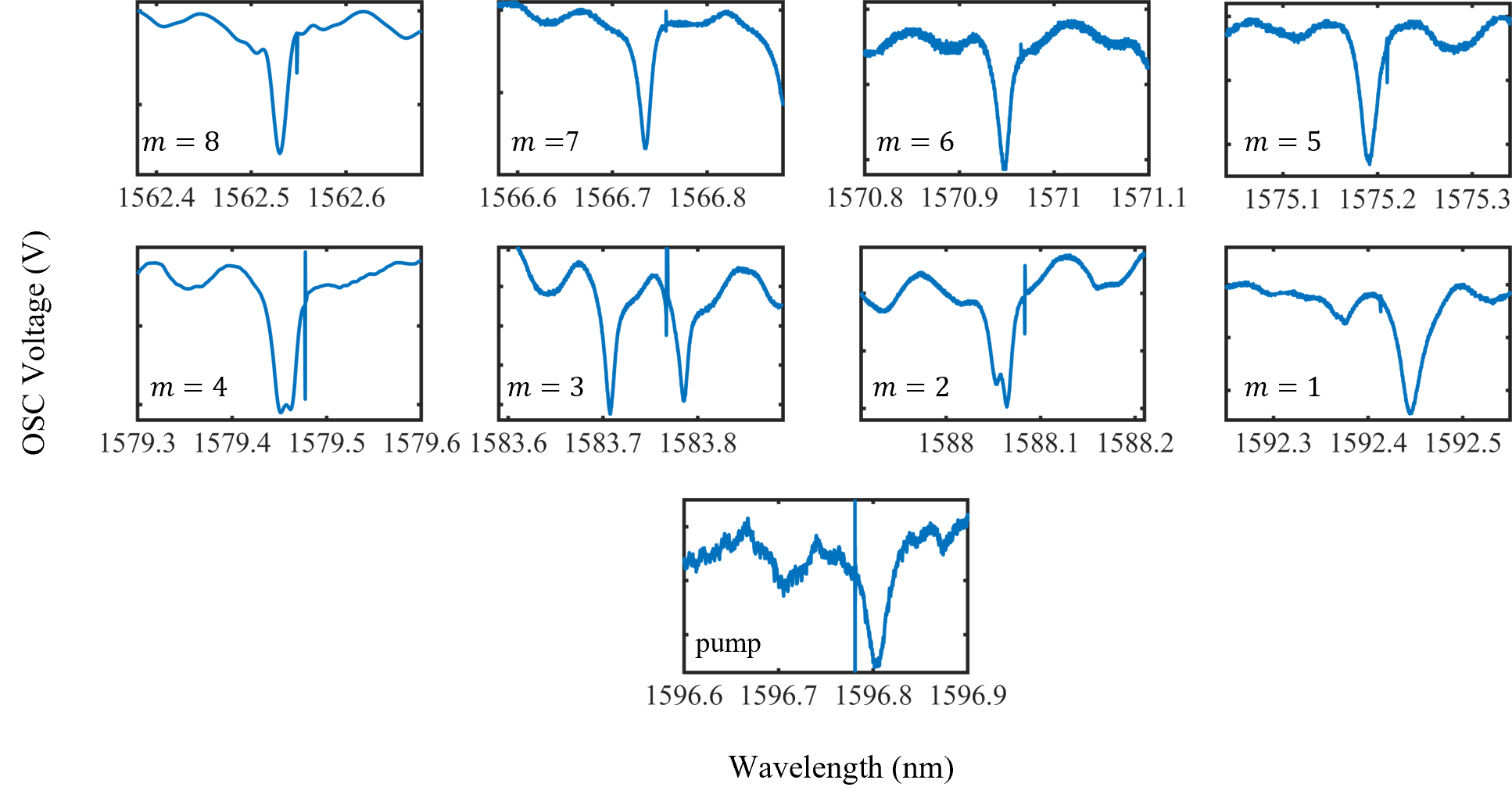}
\caption{ [Extended Data] Detuning measurement for the pump and comb lines in single DKS state. In the DKS state, the pump resides on the blue side of the partner mode resonance. Most measured comb lines are red-detuned from their corresponding
resonances, with local exceptions ($m=1,3$) associated with additional avoided-mode crossings.}  
\label{ExFig2_combpumpdetuning} 
\end{figure*}

\begin{figure*}[b]
\includegraphics[width=\textwidth]{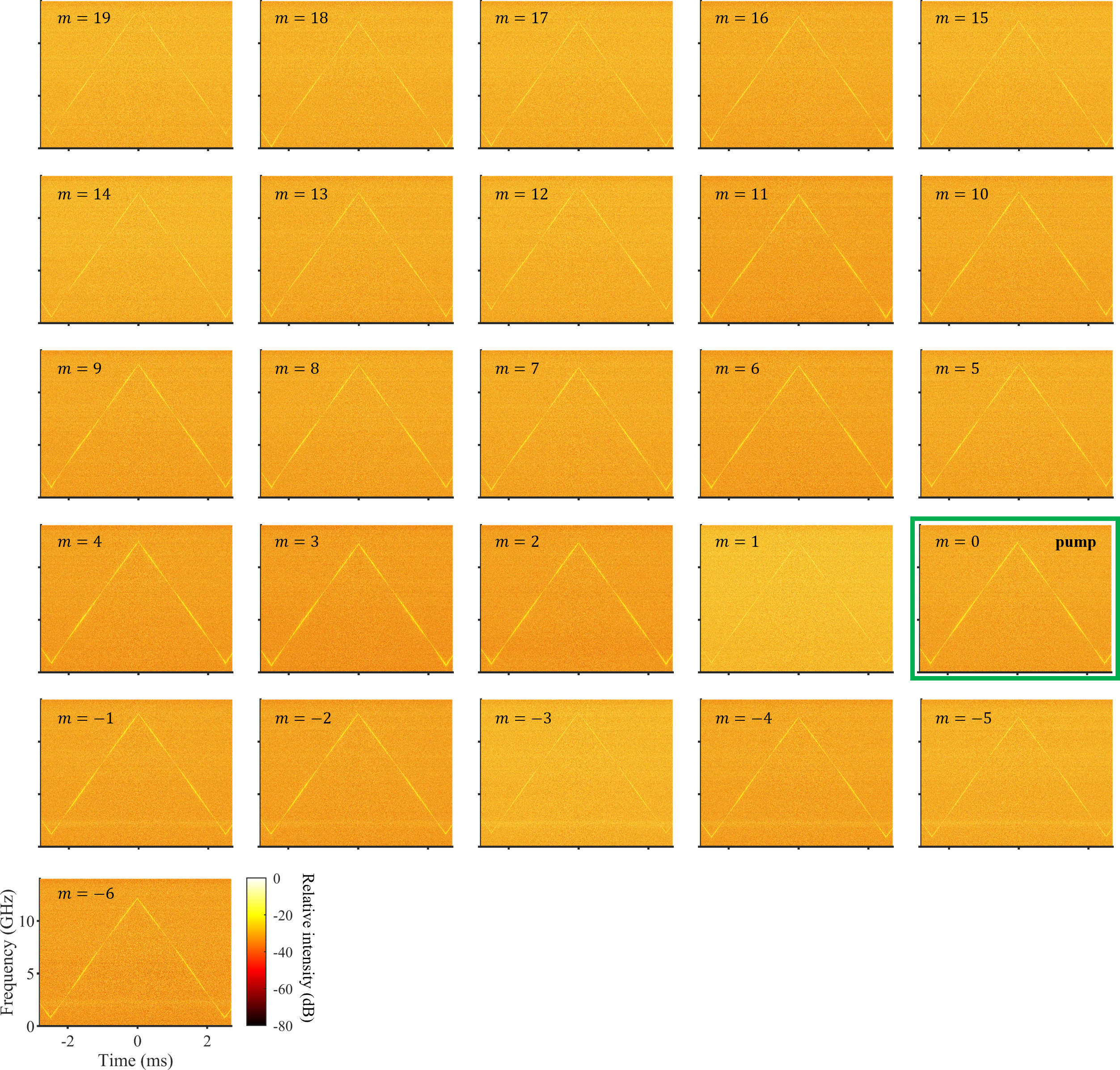}
\caption{ [Extended Data] Chirp characterization of the measured 25 comb channels and the pump. The chirping-frequency-time maps are obtained by performing short-time fast Fourier transform on the detected beating signal for each comb channel.}  
\label{ExFig4_frequency-time-map} 
\end{figure*}

\end{document}